# PROCESS ISSUES FOR A MULTI-LAYER MICROELECTROFLUIDIC PLATFORM

*S.H. Ng[1], Z.F. Wang[1], R.T. Tjeung[1] and N.F. de Rooij[2]*

[1]Singapore Institute of Manufacturing Technology
[2]Institute of Microtechnology, University of Neuchatel, Switzerland

## ABSTRACT

We report on the development of some process capabilities for a polymer-based, multi-layer microelectrofluidic platform, namely: the hot embossing process, metallization on polymer and polymer bonding. Hot embossing experiments were conducted to look at the effects of load applied, embossing temperature and embossing time on the fidelity of line arrays representing micro channels. The results revealed that the embossing temperature is a more sensitive parameter than the others due to its large effect on the polymer material's viscoelastic properties. Dynamic mechanical analysis (DMA) on polymethyl methacrylate (PMMA) revealed a steep glass transition over a 20 $^{\circ}$C range, with the material losing more than 95 % of its storage modulus. The data explained the hot embossing results which showed large change in the embossed channel dimensions when the temperature is within the glass transition range. It was demonstrated that the micro-printing of silver epoxy is a possible low-cost technique in the mass production of disposable lab chips. An interconnecting network of electrical traces was fabricated in the form of a four-layer PMMA-based device. A four PMMA layer device with interconnecting microfluidic channels was also fabricated and tested.

## 1. INTRODUCTION

Microfluidic devices can be classified into limited-function lab-on-chip (LOC) and micro total analysis system (µTAS). Most LOCs are single function and single layer devices such as mixers, separation channels, etc. Micro total analysis systems, on the other hand, are more complicated and can perform many functions such as mixing, reaction, separation, etc on a single chip. It can also consist of active valves and pumps to direct flow of fluids and metal electrodes for detection. Earlier microfluidic devices are mostly made of silicon, silicon oxide or glass partly due to the adoption of the micro-electro-mechanical systems (MEMS) fabrication processes [1]. Structural integrity and dimensional control are very good when using these materials and processes. Glass is also popular because its capillary and electroosmotic effects with water are strong. However, as the demand for microfluidic devices increases, cost becomes an issue. Cheaper materials such as polymer and mass production replication technologies would have to be implemented, especially for single-use devices. In addition, the trend would be towards cheap, disposable microfluidic devices to eliminate contamination issues. Hence, there has been a huge interest in the fabrication of microfluidic devices out of polymers in recent years [2]. The complexity of the devices has also been increasing to include on chip detection and active components such as micro-pumps, as in seen in µTAS. This would require some sort of electrical components on the devices, rather than a purely fluidic-based device such as a micro-mixer. At present, most of the devices are single layer devices with all components on a single layer of substrate. As the demand for the performance of the devices increases, multi-layer devices consisting of fluidic channels and electrical circuits would be required.

## 2. HOT EMBOSSING

In this research, the creation of micro channels on PMMA was achieved through the hot embossing process. This is a low-cost process suitable for mass production. We attempt to look at the process through a parametric study and material characterization of the polymer material.

### 2.1. Polymer flow: thick film versus thin film

In hot embossing lithography (HEL), also called nanoimprint lithography (NIL), two common applications are: thin film patterning as an alternative to photolithography, and the creation of micro channels for microfluidic devices. In the first case, a thin polymer film with a thickness typically a few hundred nanometers, is spin coated on a semiconductor wafer. After baking the polymer which is commonly a photoresist, the pattern on the stamp is transferred to the polymer through a hot embossing process. The residual layer which remains is then removed by plasma etching. The process can even replicate features down to a ten nanometers [3]. This process has great potential to replace conventional photolithography due to its simplicity and ability to replicate nano features. However, there are some issues such as its embossing uniformity, to be addressed before it can be implemented in the industry. In the second case, the target is a thick polymer substrate normally around one millimetre in thickness. The hot embossing process is used because it is simple, fast and can produce high fidelity features such as the micro channels for microfluidic devices. In such application, the feature dimensions (channel width and depth) are normally in the tens of microns.

One popular model for the hot embossing process is the squeeze film effect from hydrodynamic theory [4]. Assuming the polymer to be exhibiting ideal fluid behaviour, a fluid pressure distribution will exist in the polymer subjected to changes in the polymer film thickness. The normal load component, $L$, for a rigid circular plate approaching a rigid plane in parallel alignment can be derived from the Reynolds equation [5]:

$$L = \frac{3\pi\eta R^4 w}{2h^3} \quad (1)$$

where $\eta$ is the absolute viscosity, $R$ is the radius of the plate, $w$ is the squeeze velocity and $h$ is the film thickness. A positive pressure is generated in the fluid film existing between the plate and the plane when the two surfaces are moving towards each other. When they are moving apart, negative fluid pressures will result and this can lead to cavitation. A higher squeeze velocity and low film thickness will result in a larger load support. This provides a very important cushioning effect in bearings especially it takes a finite time to squeeze the oil from the gap. The Reynolds equation is derived from the Navier-Stokes and continuity equations for application in the hydrodynamic lubrication theory. One of the key assumptions in the derivation of the Reynolds equation is that the fluid film thickness is much smaller than the other dimensions such as the width and diameter of a journal bearing. The film thickness, in practice, is typically 3 orders of magnitude smaller than the journal dimensions. Hence, in the order-of-magnitude analysis to derive the Reynolds equation, some terms in the Navier-Stokes equations are neglected. In a further derivation of Equation (1) from the Reynolds equation, the film thickness is assumed to be "thin" so that the fluid pressure can be assumed to be constant across it.

With this in mind, it is possible to look at the situation happening in the hot embossing process. Considering a typical 100 mm diameter wafer stamp pressing against a polymer, spin coated polymer film thickness can be in the range from tens of nanometers to tens of microns. Typical bulk polymer sheets have thickness from hundreds of microns to a couple of millimeters. In any case, the wafer diameter is many times the thickness of the polymer. Hence, at the wafer level, there is no violation of the thin film assumption when the squeeze film equation is applied to the hot embossing process.

However, at the feature level, while the squeeze film effect can be applied to the case where a thin film is embossed, it cannot be applied to the case where a thick film is embossed. Fig. 1 shows the difference in the two cases. The stamp feature dimensions can range from the nanometers to microns. Hence, the squeeze film effect can be applied to the situation in Fig. 1(a) if the spin coated polymer film thickness, $h$, is many times smaller than the feature length. Laminar flow of material throughout the polymer thickness will occur between every feature and the silicon substrate. Material squeezed out between two adjacent features would be forced into the recess between them. In Fig. 1(b), the film thickness, $h$, is typically many times the feature length. Hence, Equation (1) will not be applicable. In addition, the flow across the film thickness will be very different from that of a Poiseuille flow as predicted by Equation (1). Assuming a 1 mm thick polymer substrate and 100 μm features, the situation would be analogous to placing the tip of a finger on the surface of a 1 m deep tank of water! Most of the material flow will occur around the feature rather than throughout the thickness.

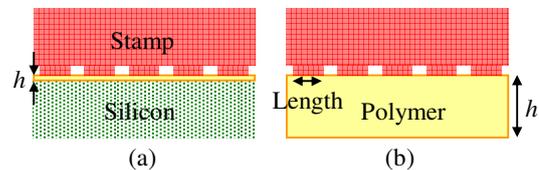

Fig. 1 Thin film (a) and thick film (b) hot embossing.

There is a fundamental difference in the polymer flow mechanism between hot embossing on thin, spin coated polymer films and hot embossing on thick polymer substrates. The mechanism in the second case should be likened to that of an indentation model (especially for well spaced out protruding features) since the contact length is much smaller than the substrate thickness. The actual situation would be much more complicated than a single indentation due to pattern dependency effects. Neighbouring patterns or features will have an effect on the embossing performance of a feature.

In both cases shown in Fig. 1, the squeeze film effect will occur at the wafer level once the polymer level hits the field region of the stamp. This will occur earlier for a stamp with recessed features than in the other extreme, a stamp with protruding features. The squeeze film effect at the wafer level can be seen when the average polymer thickness across the wafer decreases with time during embossing.

## 2.2. Experiments

Hot embossing was conducted on 1 mm thick PMMA sheets using a hybrid stamp based on the FR4 printed circuit board material. The hybrid stamp consisted of a 1.6 mm thick FR4 glass fibre reinforced epoxy topped with a 34 μm layer of copper. A pattern of raised structures from one of these copper surfaces was created by nickel electrolytic plating. The feature in this study was a channel array of about 70 μm width with 100 μm pitch. Figure 2 shows a typical profilometric scan across the array on the stamp as well as on an embossed PMMA. Channel depth was uniform with values of about 10 μm. The walls of the stamp features were not vertical but have a slight relief angle that aided stamp release after embossing. The stamp used in the experiments was measured to be 55 mm x 48 mm x 1.6 mm.

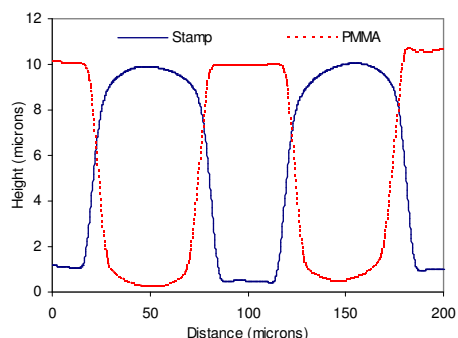

Fig. 2 Profilometric scans.

As illustrated in Fig. 3, the process consisted of a stamp or mould (with positive and negative relief structures) that was pressed against a polymer substrate that has been heated to a temperature that was normally above the glass transition temperature of the polymer material. A holding time was necessary under heat and pressure to allow the polymer to flow completely to form the patterns. Upon cooling down and separation (demoulding step) of the stamp and polymer substrate, a reversed image of the stamp pattern would be imprinted on the polymer surface. The hot embossing machine used was a bench-top hydraulic press with a capacity of 15 tonnes load. It was fitted with two 100 mm diameter heated platens (top and bottom) and these were connected to a temperature controller. A thermocouple was embedded into the top platen. Cooling was achieved through natural air cooling and through water cooled plates attached to the platens. A chiller circulated water through the two cooling plates.

Three sets of experiments were conducted to look at the effects of load, temperature and holding time on the fidelity of the embossed channels. Table 1 shows the three sets of experimental runs.

## 3. MULTI-LAYER MICROELECTROFLUIDIC PLATFORM

One of the most straight forward approaches to fabricate a three-dimensional network of embedded fluidic channels will be to create two-dimensional networks of the channels and laminate the layers together. Inter-layer interconnecting channels will have to be created also. The most challenging step in this method is the bonding process, especially for polymer-based devices. While techniques for the bonding of two flat surfaces of polymers have been established, e.g. adhesive bonding and thermal bonding, bonding two structured polymer surfaces is a much trickier business. A good bonding of surfaces containing micro channels and metal traces would be characterized by 1) minimal number of trapped air bubbles, 2) complete flow of polymer around metal traces, 3) hermetic sealing of fluidic channels, 4) minimal blockage of channels and maintaining channel dimensions. With standard thermal bonding process, the microstructures deform easily, clogging the channels since temperatures and pressures are high in order for bonding to occur. Hence, plasma or X-ray assisted thermal bonding processes have been developed. In the case of the plasma assisted process [6], both surfaces of the polymers are activated by plasma and at the same time become more hydrophilic. Thermal bonding can then be achieved with lower temperature and pressure, reducing the risks of channel deformation or even worse, channel blockage. Another popular method is the use of adhesives to bond two surfaces of polymers together. B. Bilenerg et al. [7] demonstrated the use of spin on solvated PMMA to bond an Su-8 surface with a Pyrex glass surface, creating embedded micro channels. The technique requires the bonding to be carried out in an evacuated environment. Precise control of the viscosity of the adhesive just before bonding and the bonding force is essential in preventing excessive adhesive flow into the channels.

We have designed and fabricated a multi-layer test device. Four PMMA layers each with hot embossed micro channels were bonded together simultaneously. Prior to bonding, inter-layer via holes were created by micro-drilling.

In addition to creating a network of interconnecting fluidic channels, an electrical network will have to be incorporated into a microelectrofluidic device. One of the ways to lay down conductive traces on polymers will be to use metal deposition through sputtering or evaporation. This can be combined with metal etching or lift-off techniques to create the patterns. For a polymer-based device, these wet processes often pose problems since organic solvents are used. These solvents can be found in the developers and photoresist strippers. Without strong chemical bonding between the deposited metal film and polymer surface, these solvents can attack the polymer causing metal delamination.

In this work, we adopted a low-cost micro-printing technique. A 4-layer electrical test device was designed. A stencil printer from MPM Corporation was used to deposit conductive traces on the PMMA sheets. The paste used was a low temperature cure silver-filled epoxy from Emerson & Cumming. Both electrical traces and inter-layer via holes were filled with this material. A layer by layer trace printing, via filling, thermal cure and bonding process was used.

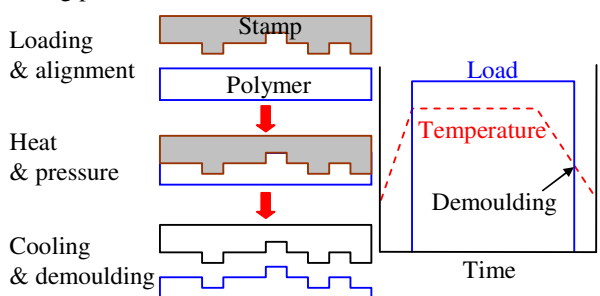

Fig. 3 The hot embossing process.

Table 1 Hot embossing runs.

| Run | Load (kg) | Time (min.) | Temperature ($^o$C) |
|---|---|---|---|
| 1 | 1200 | 10 | 130 |
| 2 | 1000 | 10 | 130 |
| 3 | 800 | 10 | 130 |
| 4 | 600 | 10 | 130 |
| 5 | 400 | 10 | 130 |
| 6 | 200 | 10 | 130 |
| 7 | 1000 | 12 | 130 |
| 8 | 1000 | 8 | 130 |
| 9 | 1000 | 5 | 130 |
| 10 | 1000 | 3 | 130 |
| 11 | 1000 | 1 | 130 |
| 12 | 1000 | 10 | 150 |
| 13 | 1000 | 10 | 120 |
| 14 | 1000 | 10 | 110 |
| 15 | 1000 | 10 | 105 |
| 16 | 1000 | 10 | 100 |

### 4. RESULTS AND DISCUSSION

Figure 4 shows some typical hot embossing results. The filling mechanism of the cavity was a single-peak mode rather than the dual-peak mode reported in some studies [8, 9]. In the dual-peak mode, the phenomenon was seen under a two-dimensional surface profilometric scan, as two peaks climbing up the cavity walls. While frictional heating and shear thinning have been proposed to be causing it, we believed another necessary condition for the dual-peak mode is the squeeze film configuration where the film thickness was much smaller than the feature size.

In our process, the polymer substrate was much thicker than the feature size. Under most conditions, the peak reached the cavity ceiling before lateral filling was completed.

We attempted to quantify three parameters to study the flow behaviour of the polymer: the maximum depth of the imprinted channels, the top and bottom widths of the channels (as illustrated in Fig. 5). In Fig. 2, it was shown that the walls of the stamp features and the embossed channels were not vertical (good for stamp release), resulting in changing widths of the channels in the vertical direction. Depths of the channels were obtained through profilometry results while the channel widths were obtained through cross sectional measurements of the samples under a microscope. Each depth data was averaged over 5 measurements while each channel width data was averaged over 3 measurements. Figure 6 shows the hot embossing results. The error bars were one standard deviation for each set of repeated measurements. The effects of embossing load and time were very minimal. The depths of the channels were almost identical under all the loads and time experimented. Channel top and bottom widths showed small decreases with both increasing load and time. The effect of embossing temperature was more pronounced. Larger decreases in channel top and bottom widths were seen with increasing temperature especially between the 100 $^o$C and 120 $^o$C range. The depth was almost constant at 10 µm except for the case at 100 $^o$C where a depth of about 3 µm was obtained.

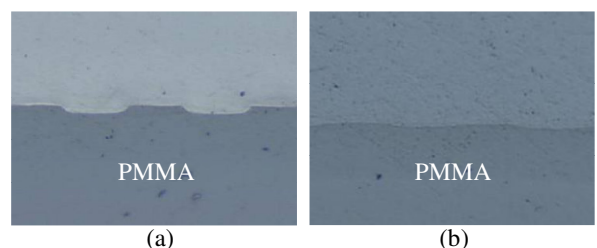

Fig. 4 Cross sectional view of embossed micro channels: (a) 1000 kg, 130 $^o$C, 10 min., (b) 1000 kg, 100 $^o$C, 10 min.

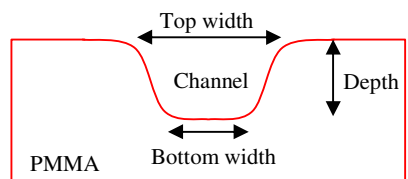

Fig. 5 Embossed feature parameters.

The PMMA was commercially available in large 1 mm thick sheets. The material was optically clear with an average surface roughness of 0.02 µm. Since the grades and properties of PMMA could vary, dynamic mechanical analysis (DMA) was conducted on the samples to determine its viscoelastic and thermal properties. The DMA experiments were conducted on a Q800 from TA Instruments using the single cantilever method. The specimen dimensions were 17.50 mm x 9.70 mm x 1.07 mm. The temperature ranges from 30 °C to 160 °C with isothermal steps of 5 minutes every 2.5 °C. A frequency sweep from 0.1 Hz to 10 Hz was carried out at each isotherm with amplitudes of 20 µm. Figure 7 shows the results from the DMA analysis. Typically, storage modulus was higher for higher frequency deformation. The glass transition temperature shifted to higher values as the frequency is increased. This could be deduced from the rightward shifts of the three plots. Tan delta values peaked well above 1.2 indicating loss modulus exceeding the storage modulus during glass transition. For the 0.1 Hz case (better estimate for the hot embossing process), the loss modulus peaked at 108 °C, indicating the material's exact glass transition temperature. The PMMA was amorphous losing more than 95 % of its storage modulus during the sharp glass transition from about 95 °C to 115 °C. For the test conducted at 0.1 Hz, the drop was from 1180 MPa at 96 °C to 36 MPa at 116 °C. This explained the results from the hot embossing tests conducted at different holding temperatures. There was a huge different in the hot embossing results for the 100 °C and 105 °C temperatures due to the rapid change in material properties in this glass transition range. Above 120 °C, there is no significant difference in the hot embossing results and this was explained by the rubbery plateau in the storage modulus after the glass transition. After glass transition, the material exhibited a gradual decrease in storage modulus (for 0.1 Hz sweep) reaching a minimum of 3.9 MPa at 138 °C. After that was another gradual increase to 6.6 MPa at 160 °C. The maximum change in storage modulus in the 130 °C to 160 °C range is about 2.5 MPa. In summary, the DMA analysis is a useful tool to understand the behaviour of polymer materials used in hot embossing experiments. The exact glass transition temperature for a polymer alone is insufficient information to explain its behaviour; rather, the glass transition range and its behaviour after that are more important. In this case, using hot embossing temperatures above 130°C (but up to 160 °C) might not necessarily improve performance or shorten embossing time due to reasons explained earlier. Hence, the number of experiments can be reduced when optimizing the process. In other cases such as semi-crystalline polymers, the situation will be different since the glass transition can happen over a much bigger range than PMMA and it exhibits different behaviour after glass transition.

Figure 8(a) shows a bonded multi-layer fluidic chip. After thermal bonding the chip was fitted with inlet/outlet connectors (supplied by Nanoport) and was connected to tubing and a syringe. Tests revealed that fluid was able to flow through all 4 layers of channels and via holes without leakage (see Fig. 8(d)). Figure 8(c) shows a cross-sectional scanning electron micrograph of a typical channel. The channel was not blocked and the overall dimensions (width and depth) were maintained. However, the shape deviated from that of a rectangular cross section due to some deformation, especially at the corners. More process control of the bonding process would improve the situation.

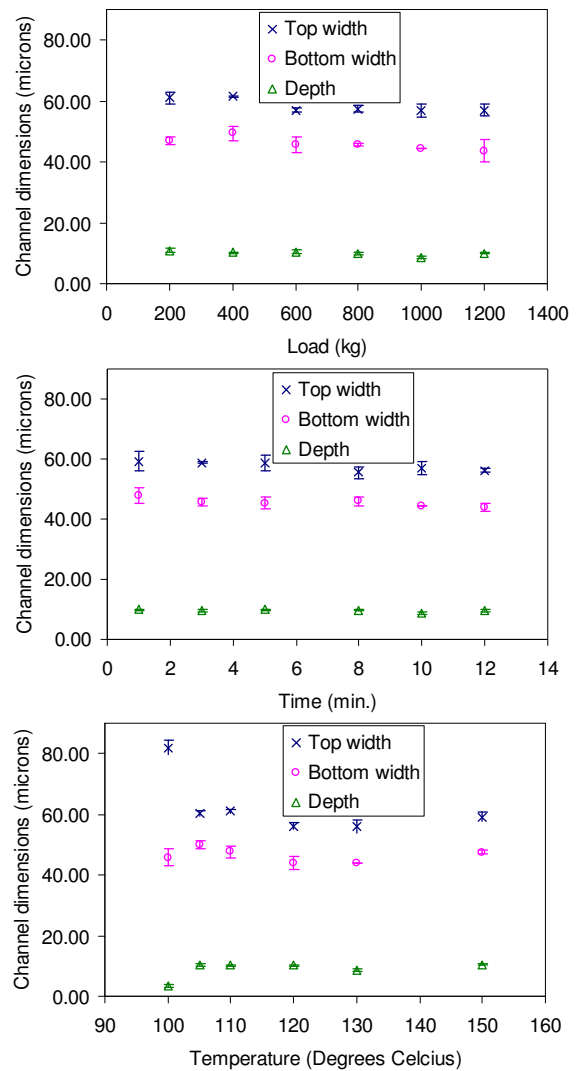

Fig. 6 Hot embossing results.

Figure 8(b) shows a bonded multi-layer electrical interconnect chip. The traces were 200 µm in width and the via holes were 1 mm in diameter. Separate printing tests using a test pattern showed that features down to 100 µm could also be printed. Electrical testing on the chip indicated connection through all the layers of the chip.

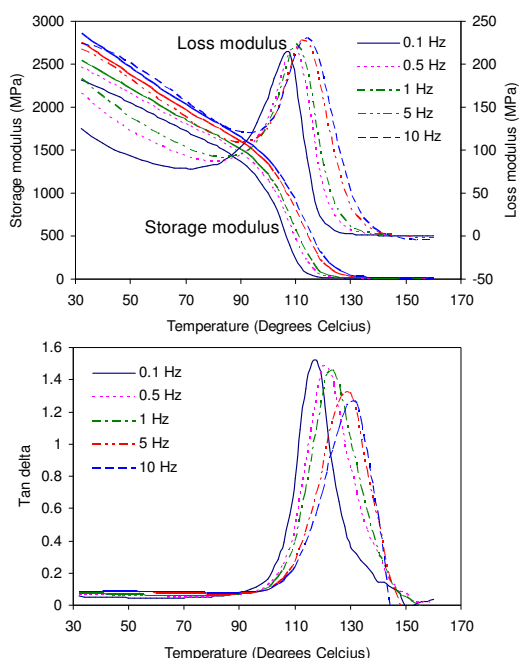

Fig. 7 DMA results.

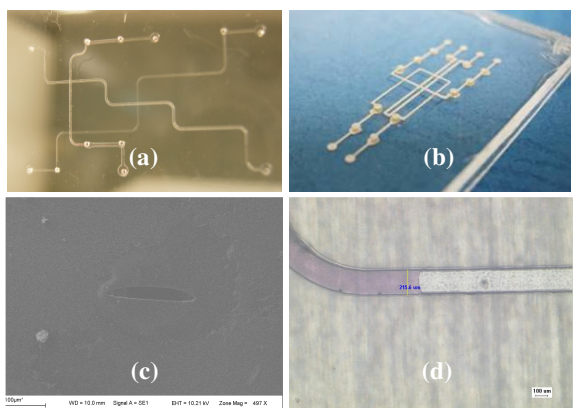

Fig. 8 Bonded multi-layer fluidic chip (a), bonded multi-layer electrical interconnect chip (b), cross-sectional scanning electron micrograph of a channel (c), fluid flow in channel (d).

## 5. CONCLUSIONS

Hot embossing experiments conducted on PMMA revealed that the embossing temperature was a more sensitive parameter compared to the applied load and embossing time. As seen in the DMA analysis on the PMMA material, a drastic drop in the storage modulus occurred in the glass transition range from ~ 95°C to 115°C. This explained the hot embossing results where large differences in channel fidelity were seen for samples embossed in the glass transition range. A four PMMA layer microfluidic test device was fabricated based on the hot embossing process and a polymer bonding technique. Fluid was able to pass through all channels without blockage and leakage. A separate four PMMA layer device consisting of interconnected traces was also fabricated using a low-cost micro-printing technique and polymer bonding. Test revealed electrical connection in the traces. The next objective in this research would be to incorporate the fluidic and electrical networks into one platform.

## ACKNOWLEDGEMENT

This research is funded by the Agency for Science, Technology and Research (A*STAR), Singapore.

## REFERENCES

[1] J. Lichtenberg, N.F. de Rooij, and E. Verpoorte, "A Microchip Electrophoresis System with Integrated In-plane Electrodes for Contactless Conductivity Detection," *Electrophoresis*, 23, pp. 3769-3780, 2002.

[2] B. Bohl, R. Steger, R. Zengerle, and P. Koltay, "Multi-layer Su-8 Lift-off Technology for Microfluidic Devices," *Journal of Micromechanics and Microengineering*, 15, pp. 1125-1130, 2005.

[3] S.Y. Chou, and P.R. Krauss, "Imprint Lithography with Sub-10 nm Feature Size and High Throughput," *Microelectronic Engineering*, 35, pp. 237-240, 1997.

[4] H.-C. Scheer, and H. Schulz, "A Contribution to the Flow Behaviour of Thin Polymer Films during Hot Embossing Lithography," *Microelectronic Engineering*, 56, pp. 311-332, 2001.

[5] B.J. Hamrock, *Fundamentals of Fluid Film Lubrication*, McGraw-Hill Inc., NJ, U.S.A, 1994.

[6] Z. Wu, N. Xanthopoulos, F. Reymond, J.S. Rossier, and H.H. Girault, "Polymer microchips bonded by $O_2$-plasma activation," *Electrophoresis*, 23, pp. 782-790, 2002.

[7] B. Bilenerg, T. Nielsen, B. Clausen, and A. Kristensen, "PMMA to Su-8 Bonding for Polymer Based Lab-on-a-chip Systems with Integrated Optics," *Journal of Micromechanics and Microengineering*, 14, pp. 814-818, 2004.

[8] L.J. Heyderman, H. Schift, C. David, J. Gobrecht, and T. Schweizer, "Flow Behaviour of Thin Polymer Films Used for Hot Embossing Lithography," *Microelectronic Engineering*, 54, pp. 229-245, 2000.

[9] H.D. Rowland, and W.P. King, "Polymer Deformation and Filling Modes during Microembossing," *Journal of Micromechanics and Microengineering*, 14, pp. 1625-1632, 2004.